\documentclass[11pt]{article}
\usepackage[margin=1in]{geometry}
\usepackage[round]{natbib}          
\usepackage{authblk}                

\usepackage{graphicx}
\usepackage{subcaption}
\usepackage{bm}
\usepackage{units}
\usepackage{xcolor}
\usepackage{newtxtext}
\usepackage{newtxmath}
\usepackage{natbib}
\usepackage{physics}
\usepackage{hyperref}
\hypersetup{
    colorlinks = true,
    urlcolor   = blue,
    citecolor  = black,
}

\newcommand{\RomanNumeralCaps}[1]
\linenumbers

\newcommand\Rey{\mbox{\textit{Re}}}
\newcommand{\uturb}{u_{\mathrm{turbulence}}}
\newcommand{\uap}{u_{\mathrm{anti\textnormal{-}puff}}}

\newcommand{\ufront}{u_{\mathrm{front}}}

\newif\ifprintcommentsShai

\newif\ifprintcommentsAnna

\newif\ifprintcommentsTODO

\printcommentsShaitrue

\printcommentsAnnatrue

\printcommentsTODOfalse

\title{Laminar gaps mirror turbulent puffs in pipe flow}

\date{}

\author[1]{Shai Kapon}
\author[2]{Tobias Grafke}
\author[1]{Anna Frishman\thanks{frishman@technion.ac.il}}
\affil[1]{Department of Physics, Technion Israel Institute of Technology, 32000 Haifa, Israel}
\affil[2]{Mathematics Institute, University of Warwick, Coventry CV4 7AL, United Kingdom}

\begin{document}
\maketitle
\begin{abstract}
Pipe flow at intermediate Reynolds numbers, between the laminar and fully turbulent regimes, takes the form of several spatially and temporally intermittent phases in which turbulent and laminar states coexist.
In the lower range, $\Rey\in (1750,2300)$, turbulence appears in the form of localized traveling structures called "puffs", which form long-lived chaotic dynamical states, whose stochastic decays and splits control the steady state intermittency. At the other end, $\Rey\in (2300,3000)$, puffs are replaced by an extended turbulent state, with laminar pockets intermittently forming and disappearing within it.
Using direct numerical simulations of pipe flow at $\Rey = 2400, 2450, 2500, 2550$, we provide evidence that these laminar gaps form a distinct dynamical state analogous to puffs: a traveling laminar pocket in a turbulent surrounding, stabilized by a shear-dependent self-tuning mechanism. 
We analyse the mean spatial profile of these gaps and show that their lifetimes are exponentially distributed, suggesting that gap closing corresponds to an escape from a chaotic saddle. Finally, we suggest these laminar gaps become unstable and disappear at a finite Reynolds number, $\Rey\sim 2900$, which can be interpreted as the onset point of spatially and temporally homogeneous turbulence. 
\end{abstract}


\section{Introduction}
\label{sec:intro}
Pipe flow is a canonical example of subcritical transition to turbulence. 
The transition is controlled by the Reynolds number, and the flow exhibits spatial and temporal intermittency with coexisting laminar and turbulent structures observed in the range  $1750 \lesssim \Rey \lesssim 3000$ \citep{rotta1956experimental}. Here,  $\Rey=D\bar{U}/\nu$ where $D$ is the pipe diameter, $\bar{U}$ is the bulk velocity defined through the cross-sectional mass flux, and $\nu$ is the kinematic viscosity.
A useful measure of this intermittency is the average turbulence fraction at steady state. As the Reynolds number is increased, the turbulence fraction in the pipe goes from zero at the lower end of the transitional range to unity at its upper end. Two basic questions then arise: How is laminar flow replaced by intermittent turbulent flow, and how does intermittent turbulent flow give way to homogeneous turbulence? 

The transition from laminar to (intermittent) turbulent flow has been at the focus of studies since the seminal experiments of~\citet{reynolds_experimental_1883}. There is now increasing evidence that this onset is a second order phase transition, belonging to the directed-percolation universality class~\citep{pomeau1986front,avila2011onset,lemoult2024directed}. The critical Reynolds number for onset is believed to be $\Rey\sim 2040$ \citep{mukund_critical_2018,lemoult2024directed}.
The basic degrees of freedom in this picture are localized, long-lived, turbulent patches called puffs, which have a characteristic structure and a well defined mean size and speed, first characterized by \citet{wygnanski1973transition,wygnanski1975transition}. In the range $Re\in (1750,2300)$ turbulence only ever occurs in the form of puffs, and it is their rates of decay and splitting that ultimately determine the turbulence fraction \citep{avila2011onset}.  

\begin{figure}
    \centering
     \includegraphics[width=1\linewidth]{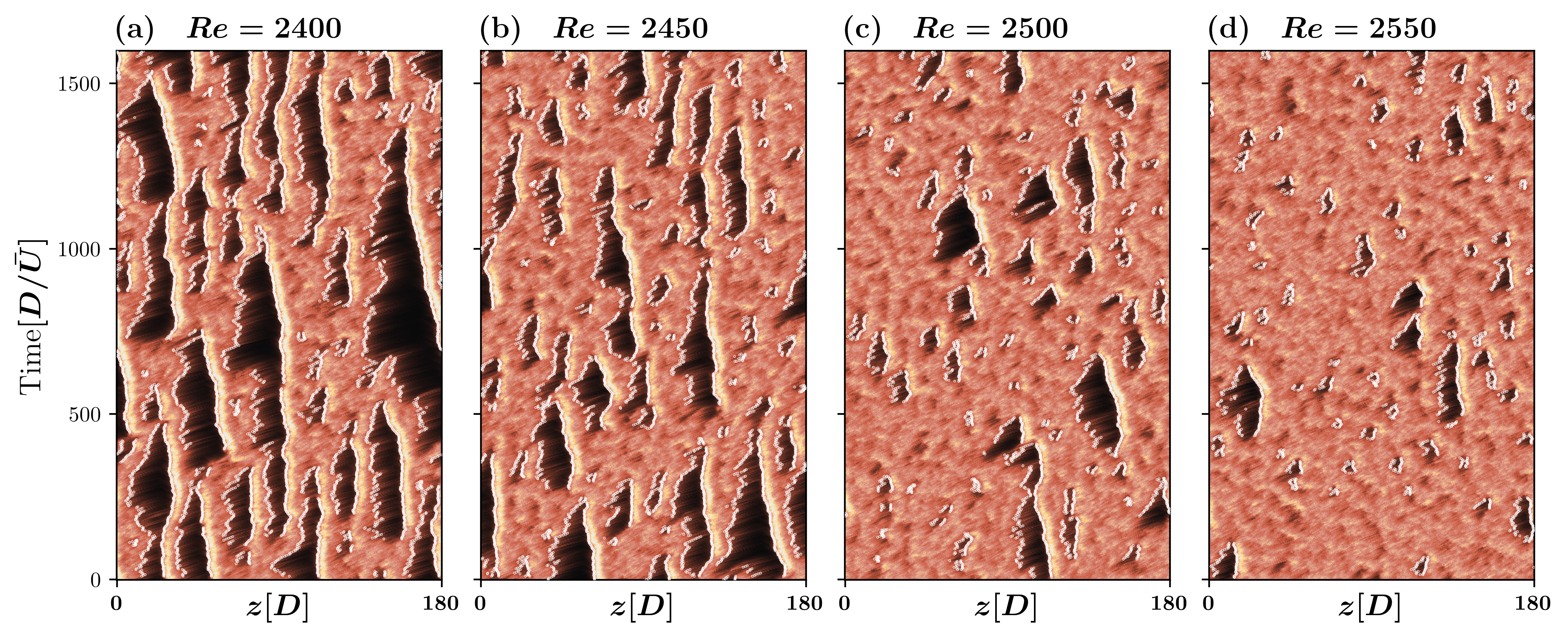}
        \caption{Space–time visualization of the flow for four different Reynolds numbers using a colour map of the turbulence level $q$ (Eq. \eqref{eq:q_def})
        with lighter colours indicating higher turbulence levels. White contours delimit the identified laminar gaps. 
        }
        \label{fig:seq_of_Re}
\end{figure}

For $\Re\gtrsim2300$ the dynamics is dominated by expanding turbulent structures called slugs, instead of puffs ~\citep{wygnanski1973transition,nishi_laminar--turbulent_2008}. However, even when turbulence occupies the entire pipe, it remains intermittent at least up to $\Rey\lesssim 2900$, with laminar gaps spontaneously opening and closing in the bulk, see Figure~\ref{fig:seq_of_Re} where black regions correspond to laminar gaps. In contrast to the puff regime, there have been relatively few studies dedicated to this intermittent phase. Qualitatively, it is observed that as the Reynolds number is increased, gaps become increasingly rare and short lived  \citep{moxey2010distinct,barkley2011simplifying,avila2013nature}. There are two possible explanations for this: Either a transition from intermittent to statistically uniform turbulence should occur at some finite Reynolds number, or rare fluctuations could perceivably open gaps at arbitrarily high Reynolds numbers~\citep{avila2013nature}, so no sharp transition exists.

Here we focus on this intermittent turbulent phase. Using direct numerical simulations (DNS) of pipe flow in the range $2400\leq\Rey \leq 2550$, we study the dynamical structures governing it. We provide evidence that (most) laminar gaps in this regime are long-lived dynamical structures mirroring the behaviour of puffs, which we term anti-puffs, in accordance with a previous suggestion based on the Barkley model~\citep{frishman2022dynamical}.
Similarly to the self-regulation process of a puff, the size of these laminar pockets is regulated by the adjustment of their (strong) downstream front to the speed of their (weak) upstream front, see Figure \ref{fig:sketch_of_fronts}(b). We further show that anti-puffs have exponentially distributed lifetimes with a mean that decreases with the Reynolds number. Such memoryless closing of the gaps hints at an escape process from a chaotic saddle. Finally, we suggest a scenario where anti-puffs become unstable and disappear at a finite Reynolds number. Considering anti-puffs as the building blocks of intermittent turbulence, such a point would correspond to the onset of statistically uniform turbulence.

\section{Laminar gaps in transitional pipe flow}
We perform DNS of pipe flow at $\Rey=2400,2450,2500,2550$ using the open-pipe open-source code \citep{openpipeflow}, with a temporal resolution of $2.5\times 10 ^{-3} \unit{D/\bar{U}}$.
We use a pipe of length $L_z = 180\unit{D}$ and a fixed spatial resolution of $(N_r, N_z, N_\theta)=(72, 2048, 48)$ corresponding to the number of modes in the radial, axial and azimuthal direction. Increasing axial resolution does not change the results.

The spatio-temporal evolution of the turbulence level with the Reynolds number is shown in Fig~\ref{fig:seq_of_Re} for our four Reynolds numbers. Here, we define the turbulence level $q(z)$ as 
the square root of the cross-sectionally averaged cross-stream energy
\begin{align}
    q(z)\equiv\left(\frac{1}{\pi 
    R^2}\int_{0}^{2\pi}\int_{0}^{R} \left(u_r^2(r,z,\theta)+u_\theta^2(r,z,\theta)\right)r\,\dd r\,\dd \theta\right)^{\frac{1}{2}}\ .
    \label{eq:q_def}
\end{align}
Lighter regions correspond to higher turbulence levels while black regions correspond to laminar gaps ($q\approx0$).
The level of turbulence is controlled by the presence of laminar gaps, and in particular their rates of formation and disappearance, gaps becoming increasingly sparse as $\Rey$ increases. 


Laminar gaps have two distinct laminar-turbulent interfaces. Following \citet{duguet_slug_2010,barkley2015rise,song2017speed}, we distinguish a weak front adjacent to streamwise-homogeneous turbulence from a strong front adjacent to laminar flow.  One might expect these fronts to be simply those of the slug, 
in which case, such gaps would monotonically contract, 
 the gap being the laminar complement of an expanding slug. While such large contracting gaps can be seen in Figure \ref{fig:seq_of_Re}, there are also small gaps that have a roughly constant width for long stretches of time, after which they close abruptly. 

\begin{figure}
    \centering
     \includegraphics[width=1\linewidth]{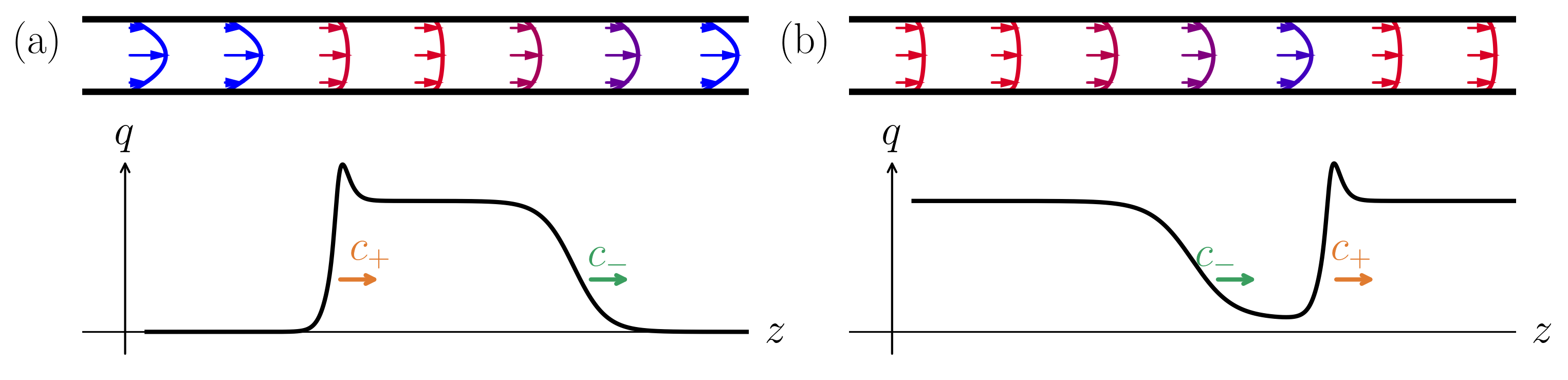}
        \caption{
        Schematic illustration of (a) a puff/slug and (b) an anti-puff. The upper panel shows the streamwise mean flow profile as a function of the streamwise coordinate $z$, and the lower panel shows the corresponding turbulence level $q$.}
    \label{fig:sketch_of_fronts}
\end{figure}

To explain how laminar gaps can persist,
 we first recall the self-regulating mechanism sustaining puffs. Puffs can also be characterized by two distinct laminar-turbulent interfaces, as sketched in Fig~\ref{fig:sketch_of_fronts} (a). The flow transitions to turbulence at
the \textit{strong upstream front}, where turbulence production is efficient, the laminar Hagen-Poiseuille parabolic profile promoting such production \citep{wygnanski1975transition,bandyopadhyay_aspects_1986,hof_eliminating_2010,song2017speed}. 
The flow then relaminarizes back at a second, turbulent-laminar, interface, the \textit{weak downstream front}. The mean flow profile is closer to a plug-like flow at this front, impeding turbulence production, so turbulence gradually decays  \citep{wygnanski1975transition,song2017speed}. Both strong and weak fronts have their own characteristic speeds \citep{lindgren1969propagation,song2017speed}, which depend on the Reynolds number and the mean flow profile at the front. We denote by $c_{+}(\Rey, \ufront)$ the strong front speed and by $c_{-}(\Rey, \ufront)$ that of the weak front, where $u$ is the centreline velocity at the front, serving as a proxy for the mean flow shear. For a puff, the profile at the strong front is fixed to be the laminar profile ($\ufront\approx 2\bar{U}$), and the size of the puff is kept approximately constant by a self-adjustment process of the downstream profile $u$ to match the speed at the upstream: $c_{+}(\Rey, 2\bar{U})=c_{-}(\Rey, \ufront)$ \citep[]{barkley2016theoretical}. 

For $\Rey>2300$, the profile at the downstream front is no longer free to adjust, pinned to the turbulent mean flow profile ($u=\uturb$). A matching of speeds for the two fronts of a puff can then no longer occur, giving rise to expanding slugs: $c_{+}(\Rey, 2\bar{U})-c_{-}(\Rey, \uturb)<0$ \citep{duguet_slug_2010,barkley2015rise}. We have previously suggested \citep{frishman2022dynamical} that laminar gaps embedded in the turbulent core can now behave similarly to turbulent puffs embedded in a laminar background: A constant gap size could be maintained if the speed of the downstream front could self-tune to that of the weak upstream front $c_{-}(\Rey,\uturb)$, see Figure~\ref{fig:sketch_of_fronts} (b). We call such persistent laminar gaps \emph{anti-puffs}. In particular, the speed of the strong downstream front could be increased by making the mean flow profile at that front sufficiently blunted, $\uturb<\ufront<2\bar{U}$, slowing down turbulence production in the upstream direction so that $c_+(\Rey,\ufront)>c_{+}(\Rey,2\bar{U})$. A choice $u=\uap$ such that the two front speeds match, $c_-(\Rey,\uturb)=c_+(\Rey,\uap)$, could then be possible, giving rise to a new type of strong front in addition to the one occurring at $\ufront\approx 2\bar{U}$. 

\begin{figure}
     \centering
    \includegraphics[width=1\linewidth]{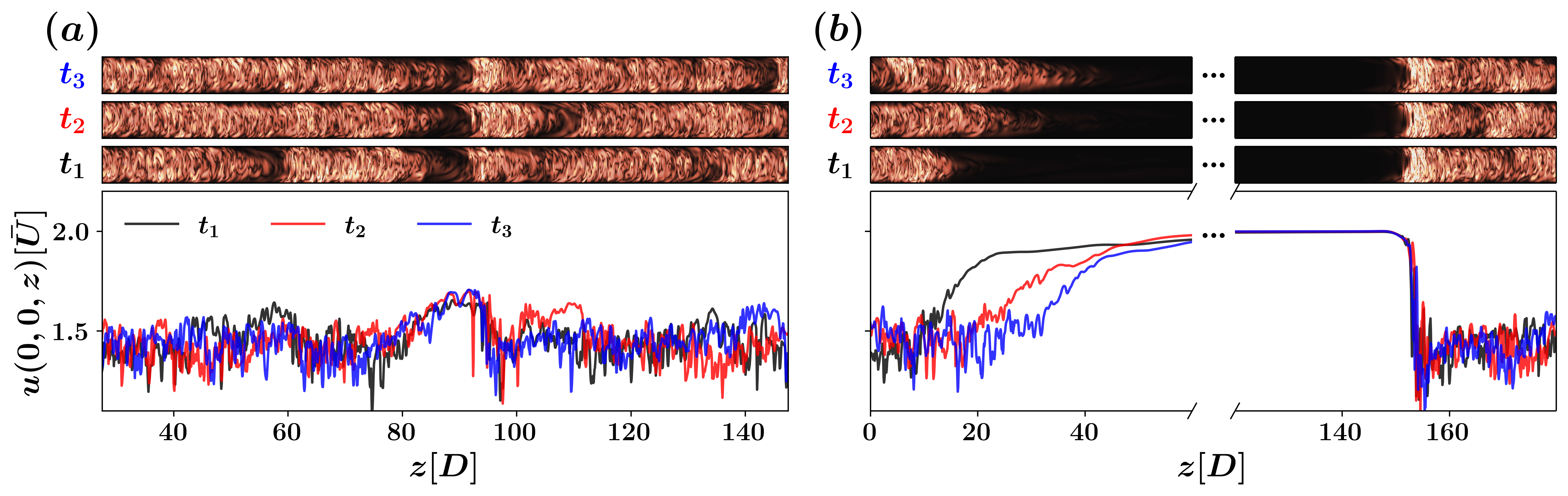}
        \caption{Comparison between an anti-puff (a) and a laminar complement of a slug (b) at $\Rey=2500$,  shown on identical spatial and colour scales. Top panels show the turbulence level (without averaging) visualized in a cross-section through the pipe at three times separated by $\Delta t = 20\unit{ D/\bar{U}}$ , plotted in a reference frame moving with the strong front. Bottom panels show the corresponding centreline velocity. For the slug, part of the laminar region between the fronts is compressed for clarity. }
        \label{fig:gap pic}
\end{figure}

To demonstrate these ideas we compare a large contracting laminar gap (with a strong front at $\ufront\approx 2\bar{U}$), Fig. \ref{fig:gap pic} (b), with a spontaneously generated small gap, Fig. \ref{fig:gap pic} (a), at $\Rey=2500$. 
In a periodic domain, a large contracting laminar gap is equivalent to a slug, and we thus use a slug generated from a localized turbulent initial condition. We show the instantaneous turbulence level in a cross-section of the pipe (without spatial averaging). In the lower panels we show the instantaneous centreline velocity $u$ along the pipe for three different times separated by $\Delta t=20\unit{D/\bar{U}}$.  
For the slug complement, the laminar core contracts by approximately $20D$ over $40\unit{ D/\bar{U}}$ and the centreline velocity is seen to reach the full Hagen-Poiseuille value $u= 2\bar{U}$ at its maximum.
For the small gap, the centreline velocity  reaches a lower value $u= 1.7\bar{U}$, and the gap shape and width remain essentially unchanged over the same time interval. Moreover, while it has a strong front exhibiting a maximum in the turbulence level, turbulence production is evidently less efficient there (compare the brightness of the downstream fronts between panels (a) and (b) in the upper panel of figure \ref{fig:gap pic}), confirming it is a different type of strong front.

           \begin{figure}
        \centering
    \includegraphics[width=1\linewidth]{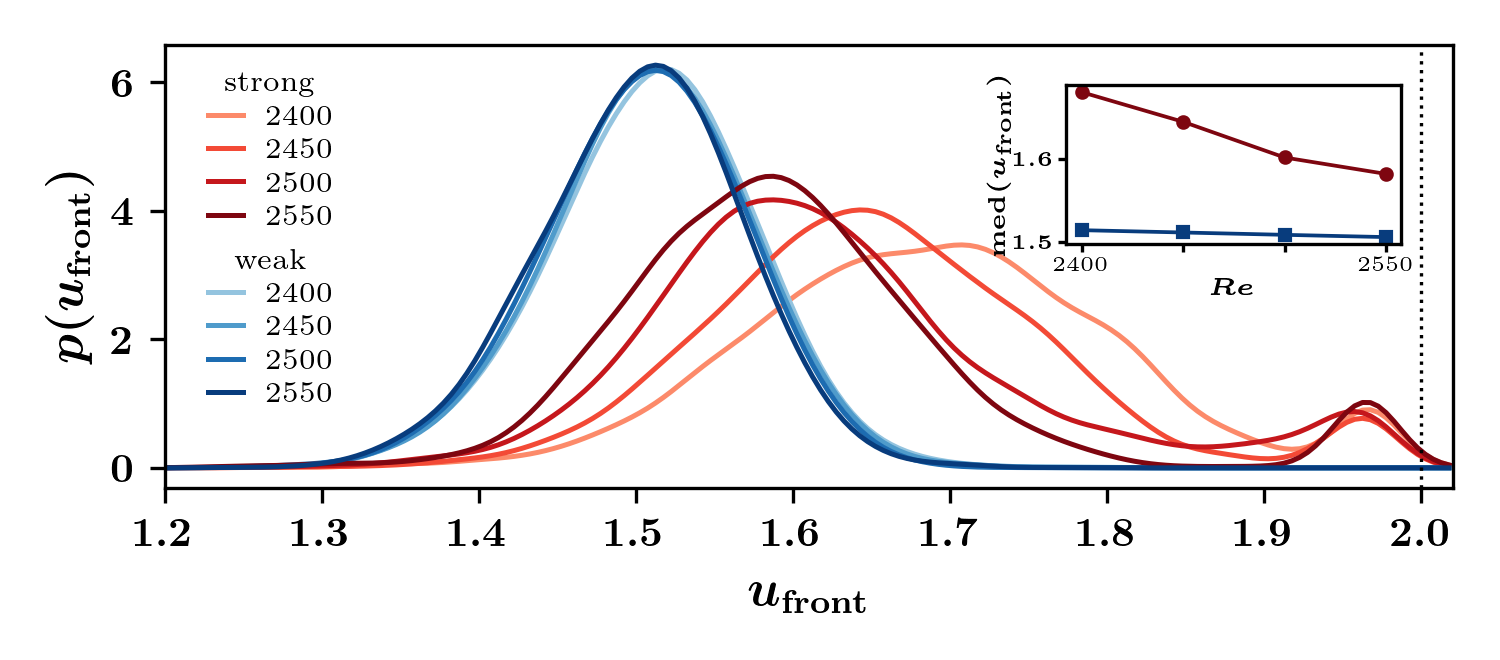}
        \caption{Distributions of the centreline velocity at laminar--turbulent fronts, $\ufront$,
          for $Re \in \{2400, 2450, 2500, 2550\}$, with strong (weak) fronts in red (blue).
          Histograms are computed with bin
          width $\Delta u = 0.005$ and smoothed with a Gaussian filter of width $4\Delta u$.
          For the strong front, the peak at $\ufront \approx 2\bar{U}$ corresponds to slugs, while the peak at lower $\ufront$ is a new type of strong front. The latter
          shifts to smaller $\ufront$ with increasing Reynolds number, whereas the weak-front
          distribution remains centred at $\ufront \approx \uturb$.
          Inset: median front velocity of the anti-puff branch,
$\mathrm{med}(\ufront)$ for $\ufront < 1.9$ for the strong front and its weak-front as a
          function of $Re$, showing the two approaching each other.}
        \label{fig:u dists}
    \end{figure} 

To analyse fronts quantitatively, we sample the fields with a sampling rate of $2.5 \unit{D/\bar{U}}$, and detect fronts by applying a threshold $q_{\mathrm{th}}=5\times10^{-2}$ to the $q$ field. We identify laminar gaps by requiring their strong front to be a long-lived ($T>30\unit{D/\bar{U}}$) connected spatio-temporal path 
(see the white contours in Fig.~\ref{fig:seq_of_Re}). The first and final three snapshots of each gap are removed from the statistics so that it only includes the mean behaviour and not the initial opening or final closing of the gap.
Varying the threshold or using $u$ instead yields qualitatively identical results. For each Reynolds number, we have included data from simulations starting from a localized turbulent spot that evolves into a long-lived slug ($T>400\unit{D/\bar{U}}$ ), and from a fully turbulent state in which laminar gaps open spontaneously. 
Most of the runs were performed in pipes of length $L=180D$, but some data from shorter pipes is also included.

The existence of a new type of Reynolds-dependent strong front is demonstrated in Figure \ref{fig:u dists}, where we measure the distribution of the centreline velocity at strong (red) and weak (blue) fronts. 
 Notably, while the distribution at the weak front is uni-modal and remains centred around $\ufront=\uturb\sim 1.5$ (blue), the strong front distribution is bi-modal, and changes with $\Rey$. The peak at $\ufront\approx2\bar{U}$ corresponds to the strong-front of a slug, while the main peak demonstrates the existence of a new type of strong front: that of the anti-puff. The change in the position of the main peak with $\Rey$ is consistent with a self-tuning process: as $\Rey$ increases the relaminarization process at the weak front slows down, requiring a more blunted profile (smaller $\ufront$) at the strong front to correspondingly slow down turbulence production to match the speed of the two fronts.

While these observations support the existence of anti-puffs, they do not explain, nor directly demonstrate, the self-tuning mechanism which keeps $\ufront\approx \uap$ on average at the downstream front, making these structures dynamically stable.     
Note that laminar gaps are essentially equivalent to the refractory tails of slugs, where the mean flow profile relaxes from the turbulent value at the upstream front, but is stopped short by an encounter with turbulent flow at their downstream end. Thus, the mean flow profile at the downstream front can be tuned by the size of the gap. A more blunted profile $\ufront<2\bar{U}$ corresponds to a small gap that does not allow full relaxation to the laminar profile. Now consider the difference between the two front speeds $c\equiv c_+(\Rey,\ufront) - c_-(\Rey,\uturb)$, measuring the widening/narrowing of the structure, exactly zero for an anti-puff $\ufront=\uap$.
Then, if a gap is initially slightly smaller than the size of an anti-puff, the laminar mean flow profile has less space to relax from the turbulent one, resulting in a flatter profile at the strong front compared to the anti-puff, $\ufront<\uap$. As a result, turbulence production at the strong front is slowed down, and the gap expands: $c>0$. Similarly, if the gap becomes wider than the anti-puff, the profile at the strong front becomes less blunted $\ufront>\uap$, turbulence production is sped up, and the structure contracts $c<0$.   
This feedback is similar to the feedback occurring for puffs \citep{barkley2016theoretical} and selects a statistically stable width at which the two front speeds are matched on average, $c \approx 0$.


\begin{figure*}
    \centering
    \includegraphics[width=1.0\linewidth]{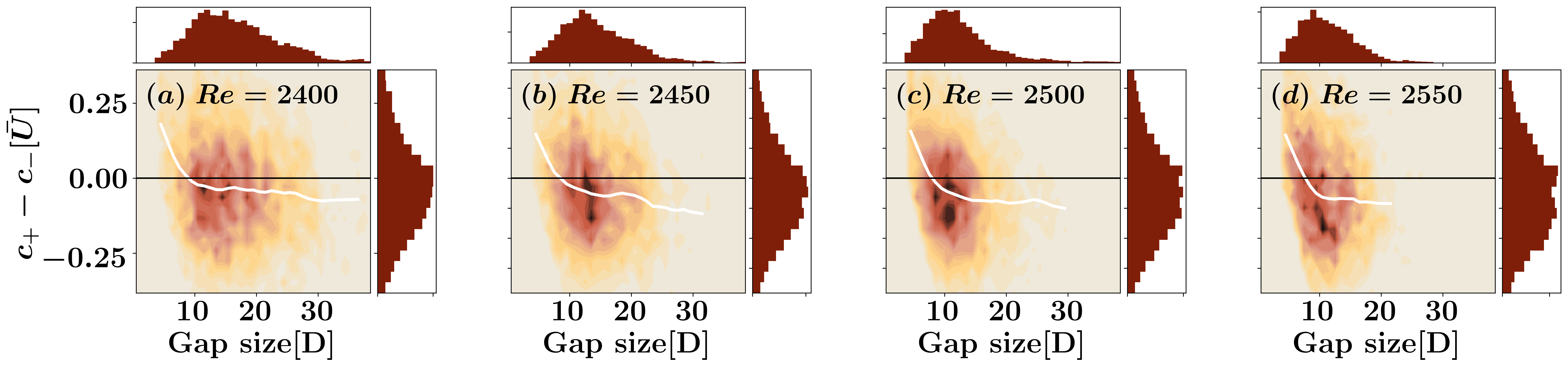}
        \caption{Two-dimensional histograms of the front-velocity difference
        $c = c_{+} - c_{-}$ against anti-puff width for (a) $\Rey=2400$,
        (b) $\Rey=2450$, (c) $\Rey=2500$ and (d) $\Rey=2550$. Positive (negative) $c$
        corresponds to expanding (contracting) gaps. The white line is the Savgol-smoothed mean over widths,
        restricted to bins containing $>25$ samples. The convergence
        to a constant at large gap sizes reflects the fixed velocity difference of slugs.
        }
    \label{fig:speed diff size corr}
\end{figure*}

To check the validity of this picture, for each instance of identified laminar gap we measure its size and the speed difference between its fronts. The data is shown as a two-dimensional histogram in Figure \ref{fig:speed diff size corr} for the four Reynolds numbers. In addition, unconditioned distributions of gap sizes and speed differences are shown on the top and right of the histogram respectively.  
 Front velocities are averaged over five consecutive snapshots.

In accordance with the above described mechanism, most gaps are concentrated around a gap size for which  $c_+ - c_-\approx 0$, where a peak in the histogram can be observed across all four Reynolds numbers in Figure \ref{fig:speed diff size corr}; for reference, a puff under the same threshold criterion is $\sim 10D$
wide. For smaller sizes, gaps tend to have a positive velocity difference $c>0$, implying their expansion, while larger gaps tend to have $c<0$ and contract. This trend is also reflected in 
the mean front speed difference $c=c_+ - c_-$, shown as a white line in Figure \ref{fig:speed diff size corr}, which crosses zero at a well-defined preferred gap size, and has a negative slope. As expected, the mean velocity difference saturates to a constant negative value for large enough gaps, corresponding to large laminar gaps where $\ufront\approx 2\bar{U}$ at the strong front. We conclude that small gaps correspond to distinct dynamical structures, leading to an operational classification of laminar gaps as anti-puffs based on their size.
 
\section{Mean profile and lifetimes of anti-puffs}

        \begin{figure}
        \centering
    \includegraphics[width=1\linewidth]{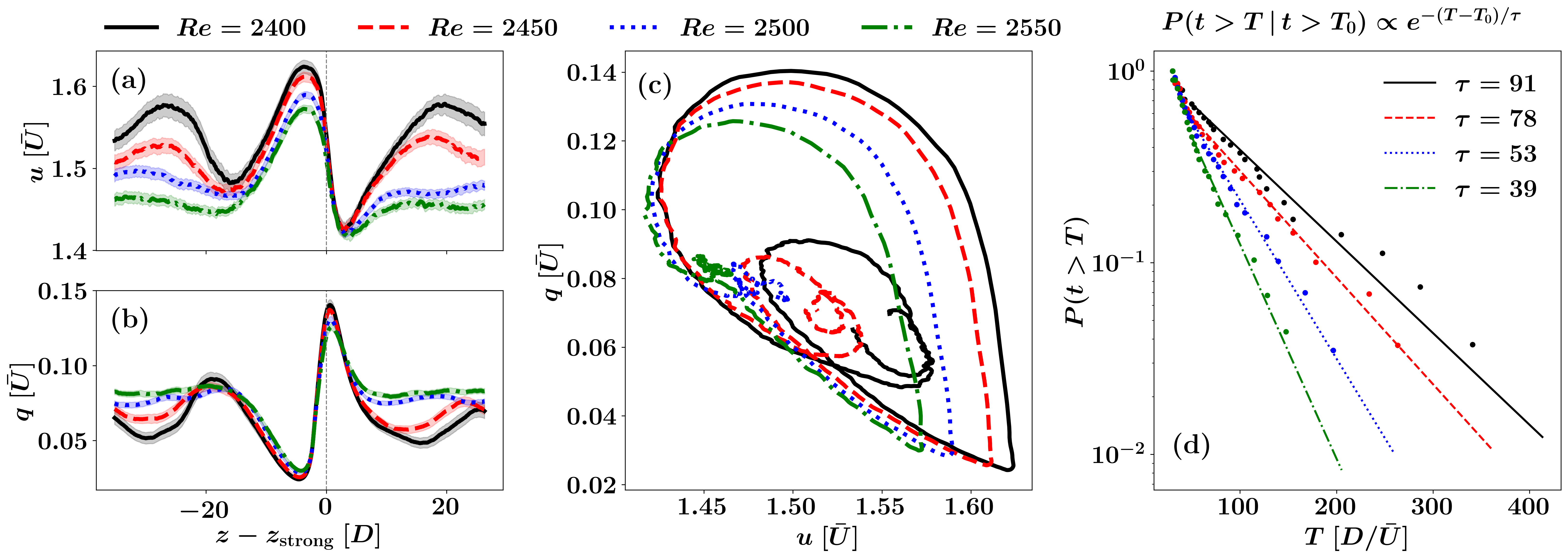}
        \caption{Mean structure and lifetime statistics of anti-puffs for $\Rey\in \{2400,2450,2500,2550\}$. Streamwise-profiles of (a) mean centreline velocity and
(b) mean turbulence level, aligned at the strong front and averaged over all identified anti-puffs. (c) Parametric plot of mean turbulence level as a function of mean centreline velocity
(d) Survival probability $P(t>T)$ for anti-puffs (complement of lifetime CDF), together with exponential fits $P(t>T | t>T_0)\propto e^{-(T-T_0)/\tau}$ starting from $T_0=50\unit{D/\bar{U}}$ and $\tau$ corresponds to the mean (fitted) lifetime
}
        \label{fig:profiles}
    \end{figure}

Having defined anti-puffs as laminar gaps smaller than a cutoff size (chosen to be $20\unit{D}$ for at least $80\%$ of their lifetime), we can now characterize their mean properties and their variation with the Reynolds number. For the statistics, 
we obtain $\sim 100 - 400$ spontaneously generated gaps for each Reynolds number in our data set.

We begin with the mean streamwise profiles of the turbulence level and the centreline velocity of anti-puffs, shown in  Figure \ref{fig:profiles} (a) and \ref{fig:profiles} (b) respectively. To obtain these profiles, individual snapshots were aligned at the strong front, identified as the local maximum in $q$. The variance across different gaps in the dataset is shown in shaded colour.
Comparing panels (a) and (b) in Figure \ref{fig:profiles} we can clearly see the negative feedback between the mean shear and the turbulence level: when the latter decreases the former increases and vice versa. To understand the dependence of the mean profiles on the Reynolds number we can use similar arguments to those for the mechanism stabilizing anti-puffs. As the Reynolds number increases, turbulence becomes more stable, slowing down the relaminarization transition at the upstream front and thus decreasing the fronts speed, so $c_{-}$ should be a decreasing function of the Reynolds number. Consequently, to match $c_{-}$, the strong front speed $c_{+}$ should also decrease, implying a more blunted profile at the strong front, with a lower maximum of the turbulence level. In particular, we expect $\uap$ to be a decreasing function of the Reynolds number. 
These features can be observed in the profiles in Figure \ref{fig:profiles} (a) and (b). They become more pronounced when the two are combined into a $q-u$ plot in Figure~\ref{fig:profiles} (c) (compare to similar plots for puffs and slugs measured by \citet{song2017speed}).     

Like puffs, anti-puffs also live for a finite time before they randomly decay, the laminar gap suddenly closing. By analogy to puffs \citep{hof2006finite}, we may expect this process to correspond to an escape from a chaotic saddle: from the anti-puff state to streamwise homogeneous turbulence. This would imply that  gap-closing is a memoryless process, and the distribution of lifetimes should follow an exponential form. This expectation is confirmed in Figure \ref{fig:profiles} (d) where we show the survival probability $P(t>T)$ of gaps 
in a log-lin plot for $\Rey\in\{2400,2450,2500,2550\}$. The data is well represented by a linear fit $P(t>T)\propto e^{-(T-T_0)/\tau}$ (lines in Figure \ref{fig:profiles} (d), starting from $T_0=50 \unit{D/\bar{U}}$)
, implying the process is Poissonian with mean lifetime $\tau$. The mean lifetimes we obtain range from $\sim90 \unit{D/\bar{U}} $ for the lowest Reynolds number to $\sim 40\unit{ D/\bar{U}}$ for the highest (the precise numerical values of the mean lifetimes are slightly sensitive to changes in the thresholds used to define anti-puffs). 
 
In addition to decays, anti-puffs are also spontaneously generated in turbulent flow, unlike puffs in laminar flow. We expect this process to also be memoryless with exponentially distributed nucleation rates, decreasing with increasing Reynolds number, as is qualitatively evident in Figure~\ref{fig:seq_of_Re}. Given the long lifetimes of anti-puffs for the lower Reynolds numbers examined here, a quantitative analysis of nucleation rates would require acquiring much more statistics, and is beyond the scope of the present work.

We would like to offer a qualitative explanation for the nucleation of anti-puffs becoming rare with Reynolds number while their lifetimes become shorter. We suggest that these processes are mediated by a critical laminar nucleus, a \textit{gap edge} state, lying on the (leaky) basin boundary between the turbulent and anti-puff states, 
similarly to the edge state between a puff and laminar flow, \citep{schneider2007turbulence,skufca2006edge}. For nucleation, as $\Rey$ increases we expect the turbulent state to become more stable and the gap-edge to become harder to reach. To explain the lifetimes, we expect the gap-edge to be a minimal gap, with the minimal possible distance between upstream and downstream fronts, having the most blunted profile possible at the downstream front. Since the profile at the strong front of an anti-puff becomes increasingly blunted with increasing $\Rey$ (quantified in the inset of Fig~\ref{fig:u dists}), anti-puffs approach the gap edge as $\Rey$ increases, decreasing their stability and hence their lifetime.  

\section{Discussion: onset of homogeneous turbulence and of puff-dominated turbulence}

Our results suggest that anti-puffs are the building blocks of intermittency for turbulence at $\Rey \gtrsim 2300$, the level of turbulence at a given $\Rey$ set by their nucleation and decay rates. The onset of homogeneous turbulence can then occur in one of two ways: a smooth crossover, where anti-puff nucleation becomes ever rarer and their lifetimes ever shorter as $\Rey$ is increased, or via a sharp transition, where anti-puffs are absent in the steady state above a critical $\Rey$. Note that a directed-percolation type transition is not to be expected here since the turbulent state is not absorbing: anti-puffs can be generated spontaneously.

We suggest that the transition to streamwise-homogeneous turbulence is sharp, occurring at $\Rey\approx 2900$ where we suggest that anti-puffs become unstable. 
The existence and stability of anti-puffs relies on the matching between the front speeds $c_-=c_+$. To see when this condition breaks, recall that turbulence in the bulk is advected with a constant speed in the lab reference frame \citep{song2017speed}, termed $c$ here. Since turbulence is \emph{produced} at the downstream strong front, for any profile $\ufront$ the front moves slower than the turbulence advection speed: $c_+<c$. We can deduce the speed of the upstream front of an anti-puff from measurements of the speed of the downstream front of a slug (as the two are equivalent). This speed was measured to be faster than that of turbulence for $\Rey>2900$ by \citet{song2017speed}, from which we can infer that $c_->c$ for the upstream front of an anti-puff at these $\Rey$. This implies that $c_+<c_-$ and an equality is impossible for all profiles $\uturb\leq\uap\leq2\bar{U}$ at the downstream front of the anti-puff. Laminar gaps therefore contract on average for $\Rey>2900$ and no stable anti-puff can exist. Note that for $\Rey\gtrsim 2900$ the downstream front of a slug is observed  by \citet{song2017speed} to randomly switch between strong and a weak fronts. Thus, the loss of stability of an anti-puff is statistical: the upstream front of laminar gaps experiences random switching of the profile $\ufront$ between $\uturb$ and $\approx2\bar{U}$, such that $c_-(\Rey,\ufront)>c$ on average.

Finally, we comment on the low Reynolds number end $\Rey\approx 2300$, around the transition point from puffs to slugs. Below it, anti-puffs are replaced by puffs. Indeed, as $\Rey$ decreases, $\uap$ increases towards $2\bar{U}$, see the main red peak in Figure \ref{fig:u dists} as compared to the isolated peak at $\ufront\approx 2\unit{\bar{U}}$. We thus expect there to be no real distinction between puffs, slugs and anti-puffs, at $\Rey\approx 2300$, all three having the same weak and strong fronts. In practice, however, there is a range of Reynolds numbers around $2250<\Rey<2350$ with jammed puffs in the steady state (and possibly jammed anti-puffs for $\Rey>2300$), see \citet{lemoult2024directed}. In the jammed state, the strong fronts of puffs do not occur exactly at $\ufront\approx 2\bar{U}$, affected by the structures upstream of them. This blurs the distinction between puffs and anti-puffs in the range  $2250<\Rey<2350$, so a sharp transition point is hard to identify.  

\section{Conclusions} \label{sec:conclusions}
We have studied laminar-turbulent intermittency in pipe flow at $2400\leq\Rey\leq2550$. We identified dynamically stable laminar gaps, which we term anti-puffs, as the key objects controlling such intermittency. We have shown that, mirroring puffs, these structures remain dynamically stable due to a self-regulated matching of speeds between their two laminar-turbulent fronts. In particular, the downstream front of these gaps forms a new type of strong front, distinct from that of a slug, characterized by a blunted mean flow profile. As the Reynolds number is increased, the profile at these strong fronts becomes exceedingly more blunted, slowing down the turbulence production process at the front to match the slower relaminarization process occurring at the upstream front of the gap. In future studies, it would be interesting to study such fronts in more detail, e.g. the underlying energy balance and mechanism of propagation speed, comparing to the strong fronts of slugs as in~\citet{song2017speed,wu_scaling_2023}. 

We further analysed the mean characteristics of anti-puffs. A key finding is that their lifetimes are exponentially distributed, the mean lifetime decreasing with Reynolds number. This suggests that the underlying gap closure mechanism is one of an escape from a chaotic saddle. We further propose that anti-puffs cease to exist above a critical Reynolds number $\Rey>2900$, above which laminar gaps always contract on average. This critical Reynolds number can thus be interpreted as the transition point from intermittent turbulence to homogeneous turbulence.

Our work suggests that laminar gaps should be considered as fundamental ingredients on the route to fully developed turbulence of wall-bounded flows. Indeed, beyond pipe flow, similar laminar gaps are observed in plane Couette flow. They appear in a transitional regime between a patterned state and homogeneous turbulence,  \citep{duguet_formation_2010,tuckerman_patterns_2011} and exhibit exponential lifetime distributions \citep{gome_patterns_2023}. On the other hand, isolated laminar gaps seem to be absent from channel flow, where there is apparently a direct transition between patterned turbulence and homogeneous turbulence \citep{tuckerman_turbulent-laminar_2014,shimizu_bifurcations_2019,kashyap_linear_2022}. Understanding the role isolated laminar gaps, or their absence, play in organizing intermittent laminar-turbulent phases and the transitions between them, is an interesting avenue for future work.
 



\bibliographystyle{plainnat}
\bibliography{jfm}




\end{document}